\documentclass[12pt]{article}
\usepackage{latexsym}

\def\bs{\begin{subequations}}
\def\es{\end{subequations}}

\catcode`\@=11

\newtoks\@stequation
\def\subequations{\refstepcounter{equation}
  \edef\@savedequation{\the\c@equation}%
  \@stequation=\expandafter{\theequation}
  \edef\@savedtheequation{\the\@stequation}
  \edef\oldtheequation{\theequation}%
  \setcounter{equation}{0}%
  \def\theequation{\oldtheequation\alph{equation}}}

\def\endsubequations{\setcounter{equation}{\@savedequation}%
  \@stequation=\expandafter{\@savedtheequation}%
  \edef\theequation{\the\@stequation}\global\@ignoretrue}

\makeatletter
        \renewcommand{\theequation}{\thesection.\arabic{equation}}%
        \@addtoreset{equation}{section}%
\makeatother

\renewcommand{\thefootnote}{\fnsymbol{footnote}}

\begin{document}

\begin{titlepage}

 November 10, 2014

\begin{center}        \hfill   \\
            \hfill     \\
                                \hfill   \\

\vskip .25in

{\large \bf Nonorthogonal bases in variational calculations and the loss of numerical accuracy \\}

\vskip 0.3in

Charles Schwartz\footnote{E-mail: schwartz@physics.berkeley.edu}

\vskip 0.15in

{\em Department of Physics,
     University of California\\
     Berkeley, California 94720}
        
\end{center}

\vskip .3in

\vfill

\begin{abstract}
The most common method for calculating accurate numerical solutions 
for complicated linear differential equations - for example, finding 
eigenvalues and eigenfunctions of 
the Schrodinger equation for many-electron atoms - is the 
variational method with some convenient basis of functions. This leads 
to a finite matrix representation of the operators involved; and 
standard numerical operations - such as Gaussian elimination - may be 
employed. When the basis functions are not orthogonal, one expects 
substantial loss of numerical accuracy in those matrix manipulations; 
and so multiple-precision arithmetic is often required for useful 
results. In this paper, for the first time, we offer a way to estimate 
the rate at which numerical cancellations will grow in severity as 
one increases the basis size. For the familiar case of using simple 
power series, $x^{n}, n<N$ as the basis instead of orthogonal polynomials, 
we predict a loss of about 2N bits or 4N bits of numerical accuracy.

\end{abstract}

\vfill

\end{titlepage}

\renewcommand{\thefootnote}{\arabic{footnote}}
\setcounter{footnote}{0}
\renewcommand{\thepage}{\arabic{page}}
\setcounter{page}{1}

\section{Introduction}
Suppose we have the linear equation 
\begin{equation}
H\psi = E \psi ,\label{a1}
\end{equation}
where H is some specified linear operator (for example, the 
Hamiltonian operator in quantum mechanics), and we want to find the 
eigenfunction(s) $\psi$ and the eigenvalue(s) $E$.  One powerful 
approximation technique is to use the variational method, where we 
represent the function $\psi$ by some finite linear superposition of 
convenient basis functions, $u_{n}(x)$. Here I use the symbol $x$ to 
represent any convenient set of coordinates in which the operator H 
may be expressed.
\begin{equation}
\psi(x) = \sum_{n=1,N} C_{n}\;u_{n}(x).\label{a2}
\end{equation}
The full operator equation (\ref{a1}) is then replaced by a finite 
matrix representation,
\begin{eqnarray}
	\sum_{m=1,N}\; H_{n,m}\;C_{m} = E\;\sum_{m=1,N}\; I_{n,m}\;C_{m}, 
	\;\;\; n=1,N; \label{a2a}\\
H_{n,m} = \int dx\;u_{n}^{*}(x)\;H\;u_{m}(x), \;\;\;\;\; I_{n,m} = 
\int dx\;u_{n}^{*}(x)\;u_{m}(x). \label{a3}
\end{eqnarray}
We would now proceed to use standard techniques of matrix 
manipulation to get this $N^{th}$ approximation to the eigenvalue E 
and the eigenvector $\psi(x)$. A systematic increase in the size N 
would be a nice way to examine the convergence of this approximation method.

People who work in this arena find that they need increasing 
numerical accuracy in their computer programs in order to get 
reliable results as they go to higher 
orders of approximation. My own work on the ground state of the Helium atom \cite{S} shows 
this in a striking manner. Using 24,099 basis functions for this 
3-dimensional problem, I needed 120 decimals of arithmetic 
accuracy in my computer programs to get 47 decimals of reliable accuracy in 
the best approximation to the eigenvalue.

It is frequently thought \cite{D} that a major 
cause of this problem is the lack of orthogonality in the basis 
functions used. But I know of no previous analysis that might give a 
quantitative estimate of the magnitude of that effect.  The present paper offers an 
attempt to do just that.

First, look at the matrix $I_{n,m}$ defined above.  If the basis 
functions were orthogonal, this would be a diagonal matrix. Look at 
the matrix $H_{n,m}$. If the basis functions were exactly the 
eigenfunctions of the operator H, then this would also be a diagonal 
matrix. But, of course, we do not know the exact eigenfunctions and 
that is why we are engaged in using approximation techniques.

Let us dwell, however, on that first question, regarding the matrix 
$I_{n,m}$, and see what we can say about the loss of accuracy we 
might expect in trying to diagonalize it by familiar matrix techniques.

Given some arbitrary set of basis functions, $u_{n}(x)$, 
we have the familiar Gram-Schmidt method for recasting them into an 
orthogonal basis, $v_{n}(x)$.

\begin{eqnarray}
&&v_{1} = u_{1} \\
&&v_{2} = u_{2} - C_{2,1}v_{1}, \;\; C_{2,1} = 
\frac{<v_{1}|u_{2}>}{<v_{1}|v_{1}>}  \\
&&v_{3} = u_{3} - C_{3,1}v_{1} - C_{3,2}v_{2} , \;\; C_{3,1} = 
\frac{<v_{1}|u_{3}>}{<v_{1}|v_{1}>}, C_{3,2} = \frac{<v_{2}|u_{3}>}
{<v_{2}|v_{2}>} \\
&&etc.
\end{eqnarray}
where the symbol $<v|u>$ means $\int dx\; v^{*}(x) u(x)$.

That system of equations looks very much like the process of Gauss 
elimination, which we would use to reduce the matrix $I_{n,m}$ to 
diagonal form.  Yes, there may well be loss of accuracy through the 
repeated subtractions in the Gram-Schmidt process; and I am guessing 
that this will be a fair model for what goes on the the matrix 
diagonalization process for the original problem in Eq. (\ref{a2a}).

Should we believe that this model problem provides a fair representation of what 
happens with the real problem, involving the matrix of H? I will guess 
that H may be regarded as just some abstract weight function in the 
integrals shown in Eq. (\ref{a3}); and the detailed models explored 
below will have various weight functions in their integrals.

The three models examined below involve a single variable, x, with 
basis functions chosen as $x^{n}, n=0,N-1$; and the weight functions 
(metrics) lead us to identify three familiar sets of orthogonal 
polynomials, named Legendre, Laguerre, Jacobi. We have formulas 
for the integral of the square of any one polynomial; and we shall 
compare that with the largest term,  involving the integral of a single 
power of x, as will occur in expanding the polynomials in their full power 
series.

\section { $x^{n}$ vs Legendre polynomials}

Legendre polynomials are orthogonal over the range $-1 \le x \le 1$;
\begin{equation}
\int_{-1}^{1}\; dx \; P_{n}(x)\; P_{m}(x) = 
\delta_{mn}\frac{2}{2n+1}. \label{b1}
\end{equation}
One familiar formula for these polynomials is,
\begin{equation}
P_{n}(x) = \frac{1}{2^{n} n! }\;( \frac{d}{dx})^{n}\; (x^{2} -1 
)^{n};\label{b2}
\end{equation}
and from this we find the leading term in the polynomial to be 
\begin{equation}
P_{n}(x) = 2^{n} x^{n}/\sqrt{\pi n} + O(x^{n-2})\label{b3}
\end{equation}
for large n.

The question we ask is this. When we try to calculate the 
orthonormality integral, Eq. (\ref{b1}), how big is the largest term 
that occurs when we use the full power series? From Eq. (\ref{b3}) we 
see that there will be a term as big as $ \int_{-1}^{1}dx 
[2^{n}x^{n}]^{2}/\pi n = 2^{2n}\; \frac{2}{2n+1}\;\frac{1}{\pi n}$.

This says that we may expect to lose 2n bits of accuracy in the calculation.

\section { $x^{n}$ vs Laguerre polynomials}

Laguerre polynomials are orthogonal over the range $0 \le x \le 
\infty$, with the weight function $x^{\alpha}e^{-x}$;
\begin{equation}
\int_{0}^{\infty}\; dx \;x^{\alpha }\;e^{-x} L_{n}^{(\alpha)}(x)\; 
L_{m}^{(\alpha)}(x) = 
\delta_{mn}\;\frac{(n+\alpha)!}{n!}. \label{c1}
\end{equation}
The generating function for these polynomials is,
\begin{equation}
(1-t)^{-\alpha -1}e^{-xt/(1-t)} = \sum_{n=0}^{\infty}t^{n} 
L_{n}^{(\alpha)}(x);\label{c2}
\end{equation}
and from this we find the leading term in the polynomial to be 
\begin{equation}
L_{n}^{(\alpha)}(x) =  (-x)^{n}/n! + O(x^{n-1})\label{c3}
\end{equation}
for large n.

 When we try to calculate the 
orthonormality integral, Eq. (\ref{c1}), how big is the largest term 
that occurs when we use the full power series? From Eq. (\ref{c3}) we 
see that there will be a term as big as $ \frac{(2n+\alpha)!}{(n!)2}$.
Comparing this with the actual normalization integral from Eq. (\ref{c1})
we see that, for large n, we may expect to lose 2n bits of accuracy in the calculation.

\section { $x^{n}$ vs Jacobi polynomials}

Jacobi polynomials are orthogonal over the range $0 \le x \le 1$ 
with the weight function $x^{\alpha}(1-x)^{\beta}$;
\begin{eqnarray}
F_{n}(\alpha,\beta,x) = 
x^{-\alpha}(1-x)^{-\beta}(\frac{-d}{dx})^{n}x^{n+\alpha}(1-x)^{n+\beta}\label{d1} \\
= \frac{(2n+\alpha+\beta)!}{(n+\alpha + \beta)!}x^{n} + O(x^{n-1}) 
\sim 2^{2n+\alpha + \beta}\; n!\; x^{n}/\sqrt{\pi n}.\label{d2}
\end{eqnarray}

\begin{equation}
\int_{0}^{1}\; dx \;x^{\alpha} (1-x)^{\beta} F_{n}(x)\; F_{m}(x) = 
\delta_{mn}\frac{n! (n+\alpha)!(n+\beta)!}{(2n+\alpha+\beta +1) 
(n+\alpha +\beta)!}. \label{d3}
\end{equation}

 When we  calculate the integral of the square of the $x^{n}$ term in 
 $F_{n}$, given in Eq. (\ref{d2}), and compare that with the 
 normalization integral, Eq. (\ref{d3}), we see that it is larger by a 
 factor $2^{4n}$. This says that we may expect to lose 4n bits of 
 accuracy in the calculation for large n.
 
\section{Discussion} 
The three model problems examined in the previous sections all give 
the answer that the likely numerical cancellation errors  
grow exponentially with N, if we start with a basis of one dimensional 
functions $ 1, x, x^{2}, x^{3}, \ldots, x^{N-1}$, and then seek to 
diagonalize the matrix $I_{n,m}$  But it is not the same exponential: we find 2N 
bits lost for the Legendre, 2N bits lost for the Laguerre, and 4N 
bits lost for the Jacobi.

Since Legendre polynomials can be written as a special case of Jacobi 
polynomials, their different results seem at first worrisome. But 
then we recognize that the Legendre polynomials have reflection 
symmetry, so that even and odd order polynomials  - as well as the 
simple powers $x^{n}$ - are automatically 
orthogonal to each other. The better phrasing of the results is 
then: For the polynomials on a finite interval (Jacobi and Legendre 
and Chebyschev, etc.) the expected loss of accuracy is 4N bits, where 
N is the number of powers kept. That means $1, x^{2}, x^{4}, \ldots 
x^{2N}$ for the Legendre and $1,x,x^{2}, x^{3}, \ldots x^{N}$ for the 
Jacobi.

For some discussion of the differing results, 2N vs 4N bits lost, see Appendix A.

I have not proven, but merely conjectured, that these rules are relevant 
to the problem of calculating the eigenvalue E of the matrix $<H-EI>$ 
constructed with such a basis (and some appropriate metric).

How do we extend this result to d-dimensional problems? This can 
become more complicated. Depending on how the basis functions are 
organized, one might predict a loss of 2N (or 4N) bits or 2dN (or 4dN) bits or 
something in between - where N is the maximum number of basis functions used in 
any one dimension. 

Let me check with the Helium calculations mentioned in the 
Introduction. The value of N (for any one coordinate) was 50; so the 2N rule predicts a loss 
of 100 bits, or 30 decimal places, while the 4N rule says 60 decimal 
places. In the actual computations I noted 
a loss of up to 120 - 47 = 73 decimal places. In this three 
dimensional problem I started with the famous Hylleraas variables, 
s,t,u, and then set up my basis functions as follows: defining 
$x=t/u$, with the range (-1,1), I used Legendre ploynomials 
$P_{n}(x)$; defining $y=u/s$, with the range (0,1), I used simple 
powers of y (as with Jacobi); for the variable s, range (0,$\infty$), I used simple 
powers with the exponential (as with Laguerre).  In addition, for that third 
coordinate, s, I used not only 
$s^{n}$ but also $s^{n} ln(s)$; and this may, in effect, double the 
size of the basis from N to 2N.  So it appears that my model 
calculations, with the 4N rule, give a pretty good ``ballpark'' estimate of what actually 
occurred.
\vskip 1cm 

\noindent{\bf Acknowledgment}

I thank Gordon Drake for some helpful conversation. 

\vskip 0.5cm
\setcounter{equation}{0}
\def\theequation{A.\arabic{equation}}
\noindent{\bf Appendix A}
\vskip 0.5cm

Here is an attempt to gain some understanding of why we found 
somewhat differing results for the rate of loss of accuracy between 
the two categories of model problems discussed in earlier sections 
of this paper.

Loosely speaking, the loss of numerical accuracy comes about because successive 
basis functions, say $x^{n}$ and $x^{n-1}$, differ rather little from 
each other and so the diagonalization of the matrix - analogized to 
the orthogonalization of the basis functions - involves taking a 
difference between two similar things. That invites loss of accuracy 
in a finite computational scheme.

Here we want to set up some measure of that closeness and apply it to 
the model problems considered above.

Let's look at the function $f = w\; x^{n}$, where $w$ is the weight 
function for a particular model problem.  We will picture this as a 
function of $x$ that peaks at some place, depending on n, and we want to see 
how that position changes with n. But we need to evaluate that ``distance'' in 
reference to some other ``distance'' implied by the function f.

Here is one way to do that. Calculate the average, $\bar{x} = \int 
dx\; x \;f/\int dx \; f$ and then define 
$\Delta = d \bar{x}/dn$. Next calculate the spread,
$\delta = \sqrt{ \bar{x^{2}} - (\bar{x})^{2}}$ and take the ratio 
$\Delta/\delta$ , at large n, as a measure of separateness of the 
neighboring basis functions.

First, for $w = x^{\alpha}\;e^{-x}$ on the interval $(0,\infty)$, we 
calculate,
\begin{eqnarray}
&& \bar{x} = (n+\alpha +1)!/(n+\alpha)! = n+\alpha +1,  \\
&& \Delta = \frac{d \bar{x}}{dn} = 1 , \\
&& \delta = \sqrt{n+\alpha +1}, \\
&& \Delta /\delta = 1/\sqrt{n+\alpha +1} \sim 1/\sqrt{n}. \label{A1}
\end{eqnarray}

Second, for $w = x ^{\alpha}(1-x)^{\beta}$ on the interval $(0,1)$, 
we calculate,
\begin{eqnarray}
&& \bar{x} = (n+\alpha +1)/(n+\alpha+\beta +2),  \\
&& \Delta = \frac{d \bar{x}}{dn} \sim  (\beta +1)/n^{2} , \\
&& \delta \sim \sqrt{\beta+1}/n, \\
&& \Delta /\delta  \sim \sqrt{\beta+1}/n.\label{A2}
\end{eqnarray}

In both cases, we see that the ratio $\Delta/\delta$ gets small at 
large n, meaning that the successive basis functions  are not far 
apart but rather close to each other, as expected. But the second 
model problem gives a ratio that is smaller than the first, 
increasingly so as n gets larger. This implies that the second model 
will experience more severe numerical cancellation errors; and this 
is what we found in Section 4 (the 4N rule) compared to Section 3 (the 
2N rule).

An alternative approach is to calculate as follows:
\begin{equation}
<u_{n}|u_{n-1}>/\sqrt{<u_{n}|u_{n}><u_{n-1}|u_{n-1}>} = cos \theta,
\end{equation}
where $\theta$ may be called the angle between the two neighboring 
basis functions in the vector space of functions.  For large n, 
$\theta$ will be small, indicating that the two functions are close. 
Doing this calculation for the two models considered above, we find 
$\theta \sim 1/\sqrt{2n}$ and $\theta \sim \sqrt{\beta+1}/2n$, in concordance with 
the results shown by the $\Delta/\delta$ method.

We also note that if we take the limit $\beta \rightarrow \infty$ in 
these caculations, 
before we consider large n,
then the second model gives the same answers as the first. This is to be 
expected, since $(1-x/\beta)^{\beta} \rightarrow e^{-x}$ in this 
limit.

\end{document}